\documentclass[11pt]{article}

\usepackage{vietnam}
\usepackage{amsmath}
\usepackage{amsfonts}
\usepackage{amssymb}
\usepackage{mathrsfs}
\usepackage{bm}

\newcommand{\Photo}{}
\newcommand{\ud}{\mathrm{d}}
\newcommand{\ui}{\mathrm{i}}
\newcommand{\calO}{\mathcal{O}}
\newcommand{\beq}{\begin{equation}}
\newcommand{\eeq}{\end{equation}}

\def\P{{\cal P}}
\def\bk{{\bf k}}
\def\F{{\cal F}}

\def\k{{\bf k}}

\begin{document}

\vspace*{4cm} 

\title{DIPOLAR DARK MATTER AND COSMOLOGY}

\author{Luc BLANCHET}
\address{$\mathcal{G}\mathbb{R}\varepsilon{\mathbb{C}}\mathcal{O}$,
  Institut d'Astrophysique de Paris -- UMR 7095 du CNRS,
  \\ Universit\'e Pierre \& Marie Curie, 98\textsuperscript{bis}
  boulevard Arago, 75014 Paris, France}

\author{David LANGLOIS} \address{Astroparticle \& Cosmologie
  (CNRS-Universit\'e Paris 7), \\ 10 rue Alice Domon et L\'eonie
  Duquet, 75205 Paris Cedex 13, France}

\author{Alexandre LE TIEC} \address{Laboratoire Univers et Th\'eories
  (LUTh), Observatoire de Paris, CNRS, \\ Universit\'e Paris Diderot,
  5 place Jules Janssen, 92190 Meudon, France}

\author{Sylvain MARSAT} \address{Maryland Center for Fundamental Physics
  \& Joint Space-Science Institute, \\ Department of Physics,
  University of Maryland, College Park, MD 20742, USA \\
  Gravitational Astrophysics Lab, NASA Goddard Space Flight Center, \\
  8800 Greenbelt Rd., Greenbelt, MD 20771, USA}

\maketitle

\abstracts{The phenomenology of the modified Newtonian dynamics (MOND)
  can be recovered from a mechanism of ``gravitational
  polarization'' of some dipolar medium playing the role of dark
  matter. We review a relativistic model of dipolar dark matter (DDM)
  within standard general relativity to describe, at some effective
  level, a fluid polarizable in a gravitational field. At first order
  in cosmological perturbation theory, this model is equivalent to the
  concordance cosmological scenario, or $\Lambda$-cold dark matter
  (CDM) model. At second order, however, the internal energy of DDM
  modifies the curvature perturbation generated by CDM.
  This correction, which depends quadratically on the dipole,
  induces a new type of non-Gaussianity in the bispectrum of the
  curvature perturbation with respect to standard CDM. Recent
  observations by the Planck satellite impose stringent constraints
  on the primordial value of the dipole field.}

\vspace{0.1cm}

\section{Introduction}

Contemporary cosmology, based on the so-called $\Lambda$-CDM cosmological model,
beautifully interprets all observations at cosmological scales, but at
the price of the introduction of a cosmological constant $\Lambda$ in
Einstein's field equations, and of an unknown form of non-relativistic,
non-baryonic matter called cold dark matter (CDM). This model
brilliantly accounts for the mass discrepancy between the dynamical
and luminous masses of clusters of galaxies, the precise measurements
of the anisotropies of the cosmic microwave background (CMB) by the
WMAP and Planck satellites, the formation and growth of large-scale
structures as seen in deep redshift and weak lensing surveys, and the
accelerated expansion as evidenced by the fainting of the light curves
of distant supernov{\ae}. However, the model $\Lambda$-CDM as it
stands faces some rather severe challenges when extrapolated, 
thanks to high resolution cosmological $N$-body simulations,
down to the smaller scales of galaxies~\cite{SaMc.02,FaMc.12}.

On the one hand, several predictions of the $\Lambda$-CDM model
are not confirmed by observations: Numerous satellites of
large galaxies obtained in $N$-body simulations remain unseen;
phase-space correlations of galaxy satellites instead of an expected
quasi-isotropic distribution; generic formation of dark matter cusps
in the center of galaxies while rotation curves of galaxies favor a
constant density profile; tidal dwarf galaxies dominated by dark
matter while they are predicted to be mostly baryonic. 

On the other hand, there are actual observations that are unpredicted
and unexplained (or not naturally explained) by $\Lambda$-CDM: Strong
correlation between the mass discrepancy (\textit{i.e.}, the presence
of dark matter) and the typical acceleration scale; surface brightness
of galaxies which is always below the Freeman limit; flat asymptotic
rotation curves of galaxies; baryonic Tully \& Fisher relation for
spirals; Faber \& Jackson law for ellipticals.

All these challenges are mysteriously solved, and sometimes with
incredible efficiency, by the empirical formula of MOND --- Milgrom's
Modified Newtonian Dynamics~\cite{Mi1.83,Mi2.83,Mi3.83}. Although it
is possible that some of these challenges will be solved within
the $\Lambda$-CDM model~\cite{Sw.al.03,Sp.al.05}, we take the view
that MOND's successes point toward a drastic modification of the
standard cosmological model at small scales. Note, however, that MOND
does not account for all the mass discrepancy at the intermediate
scale of galaxy clusters~\cite{An.al.08}.

A number of relativistic field theories have been proposed, recovering
the MOND formula in the non-relativistic limit, and in which dark
matter appears to be an apparent reflection of a fundamental
modification of gravity. The Tensor-Vector-Scalar (TeVeS) theory of
Bekenstein \& Sanders~\cite{Sa.97,Be.04,Sa.05} extends general
relativity with a timelike vector field and one scalar
field. Einstein-{\ae}ther theories, which involve a unit timelike
vector field that is non-minimally coupled to the metric, provide
interesting examples of relativistic MOND theories, when modified to
allow for non-canonical kinetic terms~\cite{Zl.al.07,Ha.al.08}. Other
proposals include a bimetric theory of gravity~\cite{Mi.09,Mi.10}, a
variant of TeVeS using a Galileon field and a Vainstein mechanism to
prevent deviations from general relativity at small
distances~\cite{Ba.al.11}, and a theory based on a preferred time
foliation labelled by the so-called Khronon scalar
field~\cite{BlMa.11}. The cosmology in several of these theories has
been extensively investigated, notably in TeVeS and non-canonical
Einstein-{\ae}ther theories~\cite{Sk.al.06,Li.al.08,Sk.08,Zu.al.10}.
It is however fair to say that all these theories have difficulties
in reproducing the CMB spectrum, even when adding a component of hot
dark matter~\cite{Sk.al.06}.

In the present paper we shall be interested in an alternative to CDM,
coined dipolar dark matter (DDM), which does reproduce the CMB
spectrum in first approximation. The physical motivation for this
model is the striking (and perhaps deep) analogy between MOND, in the
non-relativistic approximation, and the electrostatics of non-linear
dielectric media~\cite{Bl1.07}. In this view the dark matter appears
to be a dipolar medium which can be polarized by the gravitational
field of ordinary baryonic matter (galaxies). In the relativistic
version~\cite{BlLe.08,BlLe.09} the DDM has the potential of
reproducing MOND at galactic scales, and is strictly equivalent to
$\Lambda$-CDM at the level of first-order cosmological perturbations
around a Friedman-Lema\^itre-Robertson-Walker (FLRW)
background~\cite{BlLe.08,BlLe.09}. Thus, the DDM behaves like ordinary
CDM at early cosmological times, and furthermore the model naturally
involves a cosmological constant.

The aim of the present contribution, which is a summary of the detailed
paper~\cite{Bl.al2.13}, is to present the DDM model at second
order in cosmological perturbations, and to show that the non-linear
dynamics of DDM starts departing from that of ordinary CDM at this
order. As a result, we shall find that DDM predicts an additional
contribution to the curvature perturbation, specifically given by the
internal energy of the DDM fluid. This extra contribution in the
evolution of the curvature perturbation will be analyzed at second
order in perturbations using the formalism of Langlois \&
Vernizzi~\cite{LaVe.05,LaVe.06}. We compute the bispectrum of the
curvature perturbation, limiting ourselves to super-Hubble scales,
and find a specific contribution to non-Gaussianity due to the
DDM. Although the amplitude of the DDM non-Gaussianity signal depends
on the \textit{a priori} unknown value of the dipole moment at early
times, recent observations of the CMB by the Planck satellite impose
stringent constraints on the primordial value of the dipole
moment. In contrast with usual models of primordial non-Gaussianities,
where the curvature perturbation is frozen on super-Hubble scales
during the standard cosmological era, we find that the amplitude of
the non-Gaussianity induced by the DDM increases with time after the
radiation-matter equality on super-Hubble scales. This distinctive
feature of the DDM model, as compared with standard CDM, could thus
provide a specific signature in the CMB and large-scale structure
probes of non-Gaussianity.

\section{Model of dipolar dark matter and dark energy}

The model of dipolar dark matter (DDM) and dark energy of Blanchet \&
Le Tiec~\cite{BlLe.08,BlLe.09} is based on a matter action in curved
spacetime, which is to be added to the standard Einstein-Hilbert
action of general relativity (without a cosmological constant), and to
the actions of all the other matter fields (such as baryons, photons,
neutrinos, etc), all described in the standard way. 

The DDM fluid in a spacetime with metric $g_{\mu\nu}$ is described by
(i) a conserved current $J^\mu = \sigma u^\mu$, such that
$\nabla_\mu J^\mu = 0$, where $u^\mu$ is the four-velocity normalized
according to $u_\mu u^\mu = - 1$ and $\sigma = {\left( - J_\mu J^\mu
\right)}^{1/2}$ is the rest mass density; and (ii) a dipole moment
vector field $\xi^\mu$, which intervenes in the dynamics only through
its projection $s^\mu \equiv (\delta^\mu_{\phantom{\mu}\nu} + u^\mu u_\nu) \,
\xi^\nu$ orthogonal to the four-velocity $u^\mu$. From this vector we
construct the \textit{polarization} field $P^\mu = \sigma s^\mu$,
\textit{i.e.}, the dipole density.\footnote{Hereafter, we
  set $G = c = 1$ and use a metric signature $+2$. Greek indices
  $\mu,\nu,\dots$ are used for spacetime coordinate components and
  Latin indices $i,j,\dots$ for the corresponding spatial indices.}
The Lagrangian describing the DDM fluid simply
reads~\cite{BlLe.08,BlLe.09}
\beq\label{L} L = - \sigma + J_\mu \dot{\xi}^\mu - V(P) \, , \eeq
where the overdot stands for the covariant derivative with respect to
proper time, $\dot{\xi}^\mu \equiv u^\nu \nabla_\nu \xi^\mu$. Notice
that $J_\mu \dot{\xi}^\mu = J_\mu \dot{s}^\mu + \nabla_\mu (J^\mu
u_\nu \xi^\nu)$, so that the Lagrangian~\eqref{L} admits an equivalent
form depending only on the orthogonal projection $s^\mu$; this shows
that the only dynamical degrees of freedom of the dipole moment
$\xi^\mu$ are those of the projection $s^\mu$, which is a
\textit{spacelike} vector.\footnote{This is in sharp contrast with
  modified gravity theories such as TeVeS and non-canonical
  Einstein-{\ae}ther theories where the fundamental vector field is
  timelike.} The first term in Eq.~\eqref{L} is the Lagrangian of a
pressureless perfect fluid, \textit{i.e.}, that of ordinary CDM, while
the second term is analogous to the coupling of the charge current to
the four-potential in electromagnetism.

The potential $V$ is a function of the norm $P = (P_\mu P^\mu)^{1/2}$ of
the polarization. Its expansion in powers of $P$ is determined, in the
weak-field limit $P \ll a_0$ only, by the requirement of recovering
the phenomenology of MOND in the non-relativistic regime. Up to third
order, it reads~\cite{BlLe.08}
\beq\label{V} V(P) = \frac{\Lambda}{8 \pi} + 2 \pi \, P^2 + \frac{16
  \pi^2}{3 a_0} \, P^3 + \mathcal{O}(P^4) \, , \eeq
where $\Lambda$ is the cosmological constant and $a_0$ is the constant
MOND acceleration scale, measured to the value~\cite{SaMc.02,FaMc.12}
$a_0 \simeq 1.2 \times 10^{-10} \, \text{m} \cdot
\text{s}^{-2}$. Appearing both in the expansion~\eqref{V}, the
constants $\Lambda$ and $a_0$ should naturally be related numerically
in this model, $\Lambda \sim a_0^2$, which happens to be in good
agreement with observations. Indeed, if we define the natural acceleration
scale associated with the cosmological constant, $a_\Lambda \equiv
\frac{1}{2\pi}\,\sqrt{\Lambda/3}$, then the current astrophysical
measurements yield $a_0 \simeq 1.3 \, a_\Lambda$. The related
numerical coincidence $a_0 \sim H_0$ was pointed out very early on
by Milgrom~\cite{Mi1.83,Mi2.83,Mi3.83}. The near agreement between
$a_0$ and $a_\Lambda$ has a natural explanation in this model,
although the exact numerical coefficient between the two acceleration
scales cannot be determined. In the strong-field (but still
non-relativistic) regime $P \gg a_0$, the potential $V$ can be
adjusted so as to recover the ordinary Newtonian limit~\cite{BlLe.08}.

Varying the Lagrangian~\eqref{L} with respect to the dipole moment
$\xi^\mu$ and the mass current $J^\mu$, one obtains a non-geodesic
equation of motion for the fluid and an equation of evolution for the
spacelike projection of the dipole, namely~\cite{BlLe.09}
\begin{subequations}
\begin{align}
	\dot{u}^\mu &= - \hat{s}^\mu \, V' \, , \label{EOM}
        \\ \dot{\Omega}_\mu &= \frac{1}{\sigma} \nabla_\mu(V - P V') -
        R_{\mu\nu\rho\sigma} \, u^\nu s^\rho u^\sigma \,
        , \label{evolution}
	\end{align}
\end{subequations}
where we introduced $\Omega_\mu \equiv \dot{s}_\mu + u_\mu \left( 1 +
2 s V' \right)$, as well as the notations $V' \equiv \ud V / \ud P$
and $\hat{s}^\mu \equiv s^\mu/ s$. The first term in the right-hand
side of~\eqref{evolution} is akin to a pressure gradient, while the
second term, which involves the Riemann tensor $R_{\mu\nu\rho\sigma}$,
is analogous to the standard coupling to curvature for the motion of a
particle with spin. The stress-energy tensor of the dipolar fluid can
be derived by varying the Lagrangian~\eqref{L} with respect to the
metric, and reads~\cite{BlLe.09}
\beq\label{T} T_{\mu\nu} = \Omega_{(\mu} J_{\nu)} - \nabla_\rho \left(
P^\rho \, u_\mu u_\nu - u^\rho P_{(\mu} u_{\nu)} \right) - \left( V -
P V' \right) g_{\mu\nu} \, , \eeq
where parenthesis around indices denote symmetrisation. The first term is monopolar,
the second one is dipolar, and the third represents a dynamical dark
energy contribution. The stress-energy tensor~\eqref{T} is conserved,
$\nabla_\nu T^{\mu\nu} = 0$, as a consequence of the equations of
motion~\eqref{EOM} and evolution~\eqref{evolution}.

Up to an hypothesis of ``weak clusterisation'' of DDM during the
cosmological evolution, which has been justified in spherical symmetry
but not in the general case, the model was shown to reproduce the
phenomenology of MOND in the non-relativistic limit~\cite{BlLe.08}, in
the sense that the Bekenstein \& Milgrom~\cite{BeMi.84} modification
of the Poisson equation is recovered. The Euclidean norm of the ordinary
gravitational field is then given by the derivative of the potential,
$g=V'(P)$, and the MOND interpolating function is related to the
potential by $\mu = 1 - 4 \pi \, \Phi(g) / g$, where $\Phi(g)$ is the
inverse function of $V'(P)$, \textit{i.e.}, is such that $P =
\Phi(g)$.

\section{Evolution of the curvature perturbation in cosmology}

In the cosmological context, we now consider perturbations around an
homogeneous and isotropic FLRW background. Since the dipole vector
would break the isotropy of the background (because it is spacelike),
it must belong to the perturbation; we thus write $s^\mu = \calO(1)$.
To second order in cosmological perturbations, the dipolar fluid can
be described using an energy density $\rho$ and a four-velocity
$v^\mu$ that obey $v_\mu v^\mu = - 1 + \calO(3)$ and $\nabla_\mu(\rho
v^\mu) = \calO(3)$, and such that the stress-energy tensor~\eqref{T}
reduces to [using the expansion~\eqref{V}]
\beq\label{T2} T_{\mu\nu} = - \frac{\Lambda}{8 \pi} \, g_{\mu\nu} +
\left( \varepsilon + p \right) v_\mu v_\nu + p \, g_{\mu\nu} +
\pi_{\mu\nu} + \calO(3) \, , \eeq
where $\varepsilon$ and $p$ are the energy density and pressure of the
matter fluid, as measured in a frame comoving with $v^\mu$. We have
$\varepsilon = \rho(1+W)$, with $W$ the specific internal energy. The
anistropic stress tensor $\pi_{\mu\nu}$ is orthogonal to the velocity
and traceless: $v^\nu \pi_{\mu\nu} = \calO(3)$ and
$g^{\mu\nu}\pi_{\mu\nu} = \calO(3)$.

The internal energy $W$, pressure $p$, and anisotropic stress tensor
$\pi_{\mu\nu}$ are second-order quantities. Therefore, to first order in
cosmological perturbations, Eq.~\eqref{T2} coincides with the
stress-energy tensor of a cosmological constant and a pressureless
perfect fluid, so that the model is indistinguishable from the
concordance cosmological model~\cite{BlLe.08,BlLe.09}.\footnote{This fact can also be
  proven~\cite{Bl.al2.13} directly at the level of the
  Lagrangian~\eqref{L}.} At second order, however, deviations from
$\Lambda$-CDM appear because of the non-zero internal energy, pressure
and anisotropic stresses of DDM. Hereafter, we will only need the
expression of the specific internal energy, which
reads~\cite{Bl.al2.13}
\beq\label{W} W = 2\pi \sigma s^2 - \frac{1}{2} \, (\mathscr{L}_v s)^2
+ \calO(3) \, , \eeq
which involves the norm squared of $\mathscr{L}_v s^\mu$, the Lie
derivative of $s^\mu$ along $v^\mu$.

The imprint of these quadratic deviations from cold dark matter on the
evolution of the curvature perturbation can be investigated using the
geometric approach of Langlois \& Vernizzi~\cite{LaVe.05,LaVe.06},
which provides a non-linear generalization of the usual
gauge-invariant quantity $\zeta$ used in linear perturbation
theory. We introduce the one-form
\beq\label{defzeta} \zeta_\mu \equiv D_\mu {\cal N} -
\frac{\dot{\cal N}}{\dot{\varepsilon}} \, D_\mu \varepsilon \, , \eeq
where the local number of e-folds ${\cal N}$ obeys $\dot{\cal N} =
\frac{1}{3} \nabla_\mu v^\mu$, and $D_\mu \equiv (\delta_\mu^{\phantom{\mu}\nu} +
v_\mu v^\nu) \nabla_\nu$ is the projected derivative orthogonal to the
four-velocity. From the projected conservation law $v_\mu\nabla_\nu
T^{\mu\nu} = 0$, one can derive an evolution equation
for $\zeta_\mu$ that is valid \textit{non-linearly} and on \textit{all
  scales}~\cite{LaVe.05,LaVe.06}. Applying this result to the model of
DDM at second order in the perturbations yields~\cite{Bl.al2.13}
\beq\label{cons_law} \mathcal{L}_v \zeta_\mu = \frac{1}{3} D_\mu \dot{W}
+ \calO(3)\, . \eeq
For a FLRW background, the spatial part of the curvature one-form
$\zeta_\mu$ reduces to the ordinary gradient of the second-order
curvature perturbation scalar $\zeta$, and the evolution
equation~\eqref{cons_law} can be readily integrated on super-Hubble
scales, yielding the remarkably simple result
\beq\label{zeta_W} \zeta = \zeta_\text{CDM} + \frac{1}{3} W +\calO(3)
\, .  \eeq
That is to say, the additional contribution to the curvature
perturbation scalar due to the non-conservation of the curvature
one-form at second order is given, on large scales, by a fraction of
the specific internal energy of the dipolar fluid.

Introducing the line element $\ud s^{2} = a^2(-\ud \eta^{2} +
\gamma_{ij} \, \ud x^i \ud x^j)$ of the FLRW background, where $\eta$
is the conformal time, as well as the spatial component of the dipole
$s^\mu = (0,\lambda^i)$, with $\lambda^i = \calO(1)$, the covariant
expression \eqref{W} becomes
\beq\label{W2} W = 2\pi \bar{\sigma} a^2 \lambda^i \lambda_i -
\frac{1}{2} \lambda^i{'} \lambda'_i + \calO(3) \, , \eeq
where $\bar{\sigma}$ is the background energy density, $\lambda_i
\equiv \gamma_{ij} \lambda^j$ and $'\equiv \partial/\partial\eta$. The
specific internal energy $W$ is quadratic in the dipole moment, which
obeys the first-order evolution equation~\cite{BlLe.08}
\beq\label{eqevol} \lambda^i{''} + \mathcal{H}\lambda^i{'} -
4\pi\bar{\sigma}a^2\lambda^i = \calO(2) \, , \eeq
where $\mathcal{H} = a'/a$ is the conformal Hubble
parameter. Considering that the background Universe is a mixture of a
radiation and a pressureless matter fluid, this equation can be solved
analytically. Introducing the variable $y \equiv a/a_\text{eq}$, with
the subscript ``eq'' refering to the time of the matter-radiation
equality, when $\bar{\rho}_\text{rad} = \bar{\sigma} \equiv
\bar{\rho}_\text{eq}$, we find
\beq \lambda^i(y,\mathbf{x}) = \left(1+\frac32 y\right)
\lambda^i_*(\mathbf{x}) \, , \eeq
where we discarded the decaying mode in the solution. We introduced a
function of space, $\lambda^i_*(\mathbf{x})$, corresponding to the
primordial value of the dipole, early in the radiation-dominated
era. Now, Eq.~\eqref{W2} simplifies into
\beq\label{W_F} W(y,\mathbf{x}) = \F(y) \, \lambda_*^2(\mathbf{x}) \,
, \quad \text{with} \quad \F = \frac{9}{16} \,\Omega^2 \left( y + 2 +
\frac{4}{3y} \right) , \eeq
where $\lambda_*^2 \equiv \gamma_{ij} \lambda_*^i \lambda_*^j$ and
$\Omega / a_\text{eq} \equiv (8 \pi \bar{\rho}_\text{eq}/3)^{1/2}$ is
the inverse characteristic timescale for the collapse of a medium of
uniform density $\bar{\rho}_\text{eq}$. At late times, we see that the
correction $W$ to the conserved CDM curvature perturbation grows
linearly with the scale factor.

\section{Statistical description of vector perturbations and non-Gaussianity}

Next, we turn to the statistical description of the vector
perturbation $\lambda^i$. The additional contribution to the curvature
perturbation, given by \eqref{W_F}, depends quadratically on the
dipole moment. Our aim is to analyze the effect of this contribution
on the curvature spectrum and bispectrum. Borrowing from previous
works on the statistical description of a vector-type
perturbation~\cite{YoSo.08,Di.al.09,Va.al.09}, we decompose the
Fourier modes of $\lambda_*^i(\mathbf{x})$ into left ($L$), right
($R$), and longitudinal ($\ell$) polarizations according to
$\lambda^i_*(\mathbf{k}) = \sum_\alpha \lambda^\alpha_*(\mathbf{k}) \,
e^{i}_{\alpha}(\hat{\mathbf{k}})$, with the polarization vectors
$e^{i}_{L} = (1,\ui,0) / \sqrt{2}$, $e^{i}_{R} = (1,-\ui,0) /
\sqrt{2}$, and $e^{i}_{\ell} = (0,0,1)$, with the $\hat{\mathbf{z}}$
direction aligned with that of the wavevector $\mathbf{k}$. In absence
of a fundamental description for the dipole field, we simply describe
each polarization $\lambda^\alpha_*$ as a Gaussian and statistically
isotropic random field. We further assume that the polarizations are
not correlated among themselves or to the other quantities entering
the problem, and in particular to $\zeta_{\text{CDM}}$. The power
spectra $P_\alpha(k)$ are then defined by the relations
\beq\label{spectrum} \langle \lambda^\alpha_*(\bk) \,
\lambda^\beta_*(\bk') \rangle \equiv - (2\pi)^3 \, \delta(\bk+\bk') \,
P_\alpha(k) \, \delta_{\alpha\beta}\, , \eeq
where $\delta$ is the three-dimensional Dirac delta
distribution. Similarly, for the two-point fonction $\langle
\lambda^i_*(\bk) \, \lambda^j_*(\bk')\rangle$, the general form of the
vector power spectrum is
\beq P_{ij}(\mathbf{k}) = T_{ij}^{\rm even}(\bk) \, P_+(k) + \ui \,
T_{ij}^{\rm odd}(\bk) \, P_-(k) + T_{ij}^{\rm long}(\bk) \, P_\ell(k)
\, , \eeq
where we introduced $T_{ij}^{\rm even} = \delta_{ij} - \hat k_i \hat
k_j$, $T_{ij}^{\rm odd} = \varepsilon_{ijk} \hat k_k$, $T_{ij}^{\rm
  long} = \hat k_i \hat k_j$, and $P_\pm = \frac12 (P_R\pm
P_L)$. Notice that if the parity invariance that is manifest in the
phenomenological Lagrangian \eqref{L} was satisfied at a more
fundamental level, then we would have $P_- = 0$. We will also use the
standard definitions for the spectrum $P_{\zeta}$ and bispectrum
$B_{\zeta}$ of the curvature perturbation, namely~\cite{KoSp.01}
\begin{subequations}
\begin{align}
	\langle \zeta(\bk) \zeta(\bk') \rangle &\equiv (2\pi)^3 \,
        \delta(\bk+\bk') \, P_\zeta(\bk, \bk') \, , \\ \langle
        \zeta(\bk) \zeta(\bk') \zeta(\bk'')\rangle &\equiv (2\pi)^3 \,
        \delta(\bk+\bk'+\bk'') \, B_\zeta(\bk, \bk', \bk'') \, .
	\end{align}
\end{subequations}

Now, we come to the evaluation of the contributions of the dipole
moment to the spectrum and bispectrum of the curvature
perturbation~\eqref{zeta_W}. Since the internal energy $W$ is
quadratic in $\lambda_*^i$, there would be no tree-level contribution
if the dipole field was to be treated as a statistical
perturbation. Therefore, in a first calculation the dipole is assumed
to have a nonzero background homogeneous value $ \bar{\lambda}^i_*$,
with an additional fluctuation $ \delta \lambda^i_*$ described as
above, such that
\beq\label{lambda*} \lambda^i_*(\mathbf{x}) = \bar{\lambda}^i_* +
\delta \lambda^i_*(\mathbf{x}) \,.  \eeq
The homogeneous part of the dipole has to be small to be consistent
with the isotropy assumption for the background; it can be interpreted
as the averaged field over our observable Universe. The presence of
$\bar{\lambda}^i_*$ then allows a tree-level computation, with the
result~\cite{Bl.al2.13}
\begin{subequations}\label{P_B}
\begin{align}
	&P^{\rm tree}_\zeta(\bk) = P_\zeta^{\rm iso}(k) \Bigl[ 1 + g
    \, \bigl( \hat{\bar{\lambda}}_*^i \hat k_i \bigr)^2 \Bigr] \, ,
  \;\; P_\zeta^{\rm iso} = P_{\zeta_{\rm CDM}} + \frac{4}{9} \F^2
  \bar{\lambda}_*^2 \, P_+ \, , \;\; g = \frac{4}{9} \F^2
  \bar{\lambda}_{*}^{2} \, \frac{P_\ell -P_+}{P_\zeta^{\rm
      iso}} \label{Pzeta} \, , \\ &B_\zeta^\text{tree}(\bk,\bk',\bk'')
  = \frac{8}{27} \F^3 \, \bar{\lambda}_*^i \bar{\lambda}_*^j \left[
    P^{ik}(\k) P^{jk}(\k') + 2~{\rm perm.} \right] , \label{Bzeta}
	\end{align}
\end{subequations}
with $\hat{\bar{\lambda}}_*^i$ the direction of the background
dipole. The spectrum is dominated by its CDM component, as
$P_{\ell,+}\ll P_{\zeta_{\text{CDM}}}$, but acquires a small
anisotropy in the direction of $\bar{\lambda}_*^i$, parametrized by
$g(k)$. In general, the shape of the bispectrum is anisotropic and
quite complicated.

In a second calculation, we treated the dipole as a perturbation only,
setting $\bar{\lambda}^i_* = 0$, assumed scale invariance, $P \propto
k^{-3}$, and performed a one-loop calculation, where logarithmic
divergent integrals are regularized by introducing a cutoff $L^{-1}$,
with $L$ a length scale that can be chosen to be the Hubble radius
today. We found that the result obtained for the spectrum and
bispectrum are consistent with the expressions above, at leading order
in the cutoff, if we evaluate the tensor $\bar{\lambda}_*^i
\bar{\lambda}_*^j$ appearing in Eq.~\eqref{Bzeta} by an average over
the length scales smaller than $L$. This shows the consistency of the
two calculations in the sense of cosmic variance.

Next, to derive some quantitative estimates, the simplifying
assumption $P_\ell = P_+$ and $P_- = 0$ can be made, such that the
bispectrum \eqref{Bzeta} takes a local form. Assuming scale
invariance, the $f_{\rm NL}$ parameter, defined as $f_{\rm NL}(\bk,
\bk', \bk'') \equiv (5/6) \, B_\zeta(\bk, \bk', \bk'')/[P_\zeta(\bk)
  P_\zeta(\bk') + \text{2 perm.}]$, reads
\beq\label{fNL} f_{\rm NL} \simeq \frac{20}{81}\F^3 \,
\bar{\lambda}_*^2 \, \frac{P_+^2}{P_\zeta^2} = \frac{45}{8192} \,
\biggl( y + 2 + \frac{4}{3y} \biggr)^3 \, \frac{H_{\rm eq}^6\, a_{\rm
    eq}^6\, \bar{\lambda}_*^2 \, \P_+^2}{\P_\zeta^2} \, , \eeq
where $H$ is the Hubble parameter and $\P \equiv P \, k^3 / (2\pi^2)$
is constant for a scale-invariant spectrum. Introducing the parameter
$x \equiv (a_\text{eq}\bar{\lambda}_*) H_\text{eq}$ comparing the
dipole proper length to the Hubble radius at the matter-radiation
equality, and assuming the existence of a number $\alpha$ of order
unity such that $\P_+ \simeq \alpha\, \bar{\lambda}_*^2$, we obtain at
the last scattering surface
\beq\label{fNL_bis} f_{\rm NL} \simeq 1.5 \times 10^{17} \, \alpha^2\,
x^6 \sim 150 \left( \frac{\overline{W}}{10^{-5}} \right)^3 , \eeq
where the second estimate relies on $\alpha \sim 1$ and
$\overline{W}\sim x^{2}$. Since a value $\overline{W}\ll 10^{-5}$ is
necessary for a consistent treatment of the isotropic background, we
find (with our simplifying assumptions) that the obtained
non-Gaussianity is compatible, although marginally, with an observable
value.
Recently, the Planck collaboration reported $f_{\rm NL} = 2.7 \pm 5.8$
($68\%$ C.L. statistical) for primordial non-Gaussianities of the
local type from the CMB temperature map \cite{Ad.al.13}. This implies
the constraint $s_\text{eq} \lesssim 1.5 \times 10^{-3} \, H_\text{eq}^{-1}$
on the physical size of the dipole at the matter-radiation equality.

An important feature of the $f_{\rm NL}$ parameter obtained here is
that its amplitude grows like the cube of the scale factor $a$. The
growing character of this additional type of non-Gaussianity would
make it a distinctive effect of DDM (with respect to CDM) to look for
by comparing measurements at different cosmological epochs, for
instance in the temperature anisotropies of the CMB and in
large-scale structures. However, to fully justify such comparison and
having a meaningful test, the present calculation should be extended
to sub-Hubble scales.

\section*{Acknowledgments}

L.B. and S.M. acknowledge partial support from the Agence Nationale de la Recherche through the Grant THALES (ANR-10-BLAN-0507-01-02). D.L. acknowledges partial support from the Agence Nationale de la Recherche through the Grant STR-COSMO (ANR-09-BLAN-0157-01).

\section*{References}

\bibliographystyle{unsrt}    
\bibliography{}

\end{document}